# A Perspective on Deep Learning for Molecular Modeling and Simulations

Jun Zhang[1,2], Yao-Kun Lei[3], Zhen Zhang[4], Junhan Chang[3], Maodong Li[1], Xu Han[3], Lijiang Yang[3], Yi Isaac Yang[1,†] and Yi Qin Gao[1,3,5,6,‡]

[1] *Institute of Systems Biology, Shenzhen Bay Laboratory, 518055 Shenzhen, China*
[2] *Department of Mathematics and Computer Science, Freie Universität Berlin, Arnimallee 6, 14195 Berlin, Germany*
[3] *Beijing National Laboratory for Molecular Sciences, College of Chemistry and Molecular Engineering, Peking University, 100871 Beijing, China.*
[4] *Department of Physics, Tangshan Normal University, 063000 Tangshan, China.*
[5] *Beijing Advanced Innovation Center for Genomics, Peking University, 100871 Beijing, China.*
[6] *Biomedical Pioneering Innovation Center, Peking University, 100871 Beijing, China.*

Corresponding authors: † (Y.I.Y) yangyi@szbl.ac.cn or ‡ (Y.Q.G) gaoyq@pku.edu.cn

## Abstract

Deep learning is transforming many areas in science, and it has great potential in modeling molecular systems. However, unlike the mature deployment of deep learning in computer vision and natural language processing, its development in molecular modeling and simulations is still at an early stage, largely because the inductive biases of molecules are completely different from those of images or texts. Footed on these differences, we first reviewed the limitations of traditional deep learning models from the perspective of molecular physics, and wrapped up some relevant technical advancement at the interface between molecular modeling and deep learning. We do not focus merely on the ever more complex neural network models, instead, we emphasize the theories and ideas behind modern deep learning. We hope that transacting these ideas into molecular modeling will create new opportunities. For this purpose, we summarized several representative applications, ranging from supervised to unsupervised and reinforcement learning, and discussed their connections with the emerging trends in deep learning. Finally, we outlook promising directions which may help address the existing issues in the current framework of deep molecular modeling.

**Abbreviations**

AEV: atomic environment vector; AIMD: ab initial molecular dynamics; ANN: artificial neural network; BPNN: Behler-Parrinello neural network; CG: coarse-grained/graining; CNN: convolutional neural network; CV: collective variable; DPMD: deep potential molecular dynamics; EBM: energy-based models; ERM: empirical risk minimization; FEL: free energy landscape; FES: free energy surface; FG: fine-grained; GAN: generative adversarial network; GCNN: graph convolutional neural network; GNN: graph neural network; GP: gradient penalty; IDM: information distilling of metastability; MC: Monte Carlo; MaxEnt: maximum entropy; MI: mutual information; MD: molecular dynamics; MLP: multi-layer perceptron; MPNN: message-passing neural network; MRF: Markov random field; PES: potential energy surface; PMF: potential of mean force; RNN: recurrent neural network; TALOS: targeted adversarial learning optimized sampling; (RE-)VIFE: (reinforced) variational inference of free energy.

# Introduction

During the past decade, burgeoning development of artificial intelligence, particularly, connectionist models or artificial neural networks (ANNs), starts to transform many aspects in science. For instance, there is a burst of attempts to exploit ANNs to address challenges in the field of molecular modeling and simulations, like the potential energy fitting,[1-2] coarse graining,[3-4] enhanced sampling[5-6] and kinetic modeling,[7-8] *etc*. Technically speaking, an ANN is a parametric model consisting of a network of simple neuron-like processing units that collectively perform complex computations.[9] ANNs are often organized into layers, including an input layer that presents the data, hidden layers that transform the data into intermediate representations, and an output layer that produces a response (e.g., a label or an action). For example, the multi-layer perceptron (MLP) is one classic realization of ANNs. In recent years, connectionist models are undergoing a resurgence in the form of *deep learning* by stacking (usually much) more than one hidden layer in ANNs and giving rise to models in relatively "deep" architectures.[10-11] Most of the state-of-the-art deep ANNs are trained upon large amount of data using the backpropagation algorithm[12] to gradually adjust their connection strengths, and have achieved impressive and even superhuman performance on many challenging tasks like computer vision and natural language processing.[10-11]

A remarkable feature of ANNs is that they are universal approximators, i.e., they can compute any functions. This is known as the *universal approximation theorem*[13] which bedrocks the connectionist models. Since a wide range of processes or tasks in daily life and science can be considered as a form of function computation, this characteristic of ANNs is the reason why they are so versatile. Mathematically speaking, ANNs are essentially a way to represent functions in a certain basis, similar to polynomials and harmonic functions. What differentiates ANNs from polynomials in machine learning is their ability to hierarchically represent and approximate basis functions on their own. Therefore, (deep) ANNs exhibit a better trade-off between flexibility (or expressivity) and tractability, and are more scalable than polynomials when dealing with high-dimensional data. From this respect, deep learning can be particularly useful in molecular modeling, for example, representing the potential energy surface (PES) of Hamiltonian systems; or improving solutions to problems bounded by certain variational principles. But before successfully exploiting deep learning to solve the problems in molecular modeling and simulations, we are facing with two major challenges.

**Challenge 1.** The model architecture suitable for molecular systems is very different from traditional deep models. Although the *universal approximation theorem* confirms the existence of a certain ANN to be a universal approximator, it does not show what architecture that ANN should take in order to be truly universal. In theory, even a MLP with one single hidden layer could be universally expressive; while practitioners of ANNs found that sophisticated architectures with more hidden layers are helpful to both the training efficiency and the test performance. In terms of mathematics, designing the network architecture is equivalent to specifying the hypothesis space for the models, and the hypothesis space should naturally be different for different types of input data. In other words, instead of regarding the ANN as a sheer "black box", a rational design of the architecture according to the input data is crucial to the performance. In machine learning, capturing the inductive biases of the data is the rule of thumb for designing the network architecture. For instance, the convolutional neural network (CNN)[14] achieves great success in computer vision because it captures the fact that the identity of a visual object should be (at least) invariant w.r.t. its location in an image; while the recurrent neural network (RNN)[15] succeeds in natural language processing provided that it is able to learn the context-dependence of language semantics. The invention of attention mechanism[16-17] and residual (or skipped) connections[18] are also good examples of architecture design inspired by certain inductive biases. However, molecular systems involve some inductive biases very different from images or texts as will be discussed in Section I. Therefore, it is not straightforward to directly apply traditional ANN model architectures (like MLP, CNN or RNN) to handle molecular systems, and the lack of proper model architecture constitutes one major challenge over deploying deep learning for molecular modeling and simulations.

**Challenge 2.** The learning objectives involved in molecular modeling are very different from traditional deep learning tasks. Taking supervised learning as example, most of the applications of deep learning are focused on classification tasks, leading to a fruitful research field called deep discriminative learning.[10] In contrast, how to solve regression problems which are much more commonly encountered in molecular modeling (where the smoothness of the target functions is of particularly importance), still remains challenging and less addressed. Moreover, many long-standing problems in molecular simulations (e.g., importance sampling and dimensionality reduction) cannot be simply solved in a supervised manner. Instead, deep unsupervised learning (particularly, generative learning)[19] may bring about new solutions to such problems. Besides, in molecular modeling and simulations, one often has to deal with dynamic datasets because the samples are generated or updated on the fly in the meantime of model optimization. This is quite different from the common settings for supervised learning where the dataset is static and prepared *a priori*. Instead, one should resort to an adaptive learning approach to tackle the floating data as in reinforcement learning or active learning. In summary, applications of deep learning in molecular modeling and simulations are beyond the scope of supervised learning, and we are required to carefully formulate the learning objective and adopt the correct learning strategy.

In this review, we will describe some recent proceedings focused on the two challenges introduced above. The rest of the paper is organized as follows: In Section I, we will first summarize some state-of-the-art network architectures that can be used to model molecular systems. From Section II to Section IV, various applications of deep learning in molecular simulations under different learning strategies will be presented. We will also inspect the limitations of the existing methods and learning frameworks, as well as outlook several promising and important directions worthy of further exploration and investigation.

## Section I. Architectures of deep models for molecular modeling and simulations

### 1. Inductive biases of molecular systems

In machine learning, inductive biases are some assumptions about the data, reflecting the internal structure, or invariance, or some other intrinsic properties of the data based on empirical knowledge. Incorporation of inductive biases allows a learning algorithm to restrict the search scope for possible solutions or to prioritize one solution over another, independent of the observed data. From the perspective of deep learning, capturing the inductive biases within the data is the core of successful deep models. Examples include the success of CNNs over images and RNNs over texts etc. Therefore, in order to develop reasonable models for molecular systems, we should first be aware of the inductive biases underlying the molecular systems, as summarized below.

(1) **Reference invariance**: Properties (e.g., the energy) of molecular systems should be invariant to the trans-rotational transform of the system's coordinates. In other words, molecular properties are determined by the relative but not the absolute positions of the particles in the absence of external field(s).

(2) **Degeneracy and symmetry**: Molecular properties should be invariant w.r.t. the permutation or exchange of identical particles (e.g., atoms, molecules or residues). Besides, many molecular properties are governed by some rules about the geometric symmetry.

(3) **Smoothness**: Changes in molecular properties should be continuous and smooth w.r.t. the changes in positions of the constituting particles.

Therefore, a deep neural network can be adopted to describe molecular systems only if it is able to preserve the above inductive biases. Apparently, traditional deep models developed in machine learning community do not meet these requirements, so novel models of special-purpose architectures are largely needed. Fortunately, some deep models have already been developed for this purpose, and we will term them as the *"deep molecular models"* hereafter. Existing deep molecular models can be roughly distinguished into two types according to their architectures: the descriptor-based models or the graph neural network (GNN)[20] based models. We surveyed some representative deep molecular models in Table 1. On top of the model architecture, choosing proper activation functions and regularization techniques is also crucial for the performance of deep molecular models. We will elaborate on these topics in the following.

### 2. Descriptor-based models

As its name suggests, the descriptor-based model builds up on "descriptors": vectors that can characterize the molecular systems. Once the descriptors are chosen, they can be directly used as input to a simple feed-forward ANN which approximates a target molecular property. The resulting ANNs are thus eligible for modeling molecular systems, provided that the descriptors are trans-rotational invariant and preserve the symmetry of the system. Let **x** stand for a molecular system consisting of $N$ particles $\{\mathbf{x}_i\}_{i=1,..,N}$ (the same symbols will be consistently re-used in this paper), the descriptor-based models first transform the $i$-th particle coordinate $\mathbf{x}_i$ into a set of descriptors $\mathbf{q}(\mathbf{x}_i)$, and exploit simple ANNs like MLPs, denoted by $f_\theta$, to predict the molecular property $A(\mathbf{x})$ based on the descriptors as summarized in Eqs. (1-2). Figure 1A shows a schematic computational graph of typical descriptor-based models.

$$\mathbf{x}_i \mapsto \mathbf{q}(\mathbf{x}_i) \tag{1}$$

$$A(\mathbf{x}) = \sum_{i=1}^{N} f_\theta(\mathbf{q}(\mathbf{x}_i)) \qquad (2)$$

In machine learning, choosing "descriptors" is also known as feature engineering. In order to predict the desired molecular property, in principle, one should find a "complete set" of descriptors determining that property, because neglecting even one of the essential components would cause prediction failure. Clearly, finding a "complete set" is generally too demanding. Nevertheless, in practice, one may still manually pre-define a large set of descriptors which may contain redundant information and yield a very high-dimensional input vector, then handle this high-dimensional descriptor via ANNs. As expected, the performance of such models will heavily rely on the choice of the descriptors.

In terms of machine learning, descriptors are one kind of "*structured input*". Like other machine learning methods dealing with structured input, the performance is usually not so much sensitive to the model architecture as to the feature engineering or pre-processing. Due to this fact, research and development of descriptor-based models mainly focused on feature engineering, that is, selecting descriptors $\mathbf{q}(\mathbf{x})$. The pioneering work along this line can be dated back to the Behler-Parrinello neural network (BPNN).[21] In BPNN, the environment of each atom is transformed by specific symmetric functions and encoded into descriptors, called atomic environment vectors (AEVs). Based on these AEV descriptors, a set of MLPs output the atomic energies, the summation of which gives the overall energy. BPNN shares parameters, that is, uses a same MLP for the same type of atom, but MLPs are separately parametrized for different types of atoms. Noteworthy, the descriptors introduced by BPNN and its following variants ANI and TensorMol,[22-23] not only include two-body terms, but also account for three-body (or angular) features. Following this line, one can even develop four-body descriptors if necessary (e.g., chirality descriptors). On top of such atom-centered symmetry descriptors, many improved or alternative ways of formulating the descriptors were proposed, to name a few, bi-spectrum components,[24] Coulomb matrix,[25] smooth overlap of atomic positions,[26] etc. In a recently developed descriptor-based model, deep potential molecular dynamics (DPMD),[2] treats each atom in a local coordinate frame, of which the rotation and translation degrees of freedom are removed. The local coordinate of each atom can be regarded as one type of AEV thus can be adopted as descriptors as in BPNN.

One nicety of these descriptor-based models, especially when compared with GNN-based models which will be discussed later, is that the computational complexity is generally linearly scaled w.r.t. the number of particles ($\mathcal{O}(N)$ where $N$ is the number of particles) once the descriptors are well formulated. Nonetheless, from the introduction above, it is not hard to follow that descriptor-based models cannot scale well with the "complexity" of the molecular system. Specifically, when the heterogeneity of the system (e.g., the type of constituting particles) increases, the number of necessary descriptors will grow rapidly. As a result, applications of descriptor-based models were mostly limited to relatively simple systems.[27] Worse still, since the suitable descriptors $\mathbf{q}(\mathbf{x})$ will inevitably vary for different systems, the descriptor-based models have limited transferability between different molecular systems, thus drawbacks their applications for more general purposes like force field development. Despite of these shortcomings, development of the descriptor-based models has furthered and enriched our understanding of how to represent molecular systems.[26, 28] The various descriptors proposed in this field have found wide applications beyond deep molecular models. For example, some descriptors can be used as good collective variables (CVs) or order parameters for the analysis or enhanced sampling of material systems.[29-30] Moreover, finding good descriptors relevant to a particular property on its own can be defined as a machine learning problem, where the manual feature engineering can be replaced by automatic representation learning, which will be an important direction for further studies.

## 3. GNN-based models

In contrast to descriptor-based models, some recently proposed deep molecular models directly take the coordinates and types of the constituting particles as the input. In other words, this kind of models is able to deal with the *"unstructured input"* according to Eq. (3),

$$A(\mathbf{x}) = f_\theta(\{z_i, \mathbf{x}_i\}) \tag{3}$$

In Eq. (3), $z_i$ and $\mathbf{x}_i$ correspond to the type and coordinates of the $i$-th particle (e.g., if the constituting particle is atom, $z_i$ can be the atomic number or nuclear charge) and the neural network $f_\theta$ takes in the entire set of the particles as the input. The symbol $\{\cdot\}$ indicates that $f_\theta$ is invariant to the re-ordering of the particles. Such kind of models treats the molecular system as a dense graph, where the constituting particles correspond to the nodes (or vertices) in a graph, and two-body interactions can be straightforwardly represented by weighted edges between nodes. Representing molecules as graphs has the advantage that it naturally captures the inductive biases for molecular systems. Similar graph-based models have been developed in machine learning community to tackle with interacting particles or agents.[31] Following this line, deep molecular models can directly borrow ideas and techniques from GNNs.[20] Among the various types of GNNs, graph convolutional neural networks (GCNNs)[32-33] and message-passing neural networks (MPNNs)[34] are of particular usefulness in modeling molecular systems. Typically, in GNN-based molecular models (Fig. 1B), each particle $i$ is represented by a node vector $\mathbf{n}_i$,

$$\mathbf{n}_i = f_{\theta_{\text{embed}}}(z_i) \tag{4}$$

During learning, each node $i$ will first gather messages $\mathbf{m}_i$ from its neighbors (denoted by $\mathcal{N}(i)$), for example, through Eq. (5),

$$\mathbf{m}_i = \sum_{j \in \mathcal{N}(i)} f_{\theta_{\text{message}}}(\{\mathbf{n}_j, |\mathbf{x}_j - \mathbf{x}_i|\}; z_i) \tag{5}$$

Then each node vector is updated based on these messages, for instance, through Eq. (6),

$$\mathbf{n}_i = \mathbf{n}_i + f_{\theta_{\text{update}}}(\mathbf{m}_i; z_i) \tag{6}$$

Equations (5-6) constitute one iteration of message-passing and node update. After repeating a fixed number of such iterations, one gets the final embedding for each node $\mathbf{n}_i$ and predicts the molecular property via Eq. (7),

$$A(\mathbf{x}) = \sum_i f_{\theta_{\text{read}}}(\mathbf{n}_i) \tag{7}$$

In summary, the overall model Eq. (3) is now decomposed into four modules: $\theta_{\text{embed}}$ for node embedding, $\theta_{\text{message}}$ for message passing, $\theta_{\text{update}}$ for node update, and $\theta_{\text{read}}$ for final readout. Guided by the general principle formulated in Eqs. (4-7), different GNN-based molecular models differ with each other in their choices of the model architecture for the four specific modules. For example, SchNet[35] along with its earlier ancestor deep tensor neural networks (DTNN)[36] is a continuous analogue to the discrete GCNNs, and performs continuous convolution over the 3D Euclidean space. To achieve this goal, SchNet first employs ANNs to generate convolutional filters based on inter-atomic distances, then adopts these filters to perform convolutional operations over the particles in the space. On the other hand, PhysNet[37] represents another class of GNN-based models which replace convolutional operations with attention mechanism.[16-17] Since attention-based models have achieved the state of art in many deep learning tasks including computer vision and natural language processing,[17, 38] it is also likely that graph attention operations[39] would further improve the performance of deep molecular models.

Note that GNN-based models consist exclusively of particle-wise operations, thus the overall model can be transferable regardless of the number and type of the particles in the system. Besides, since the GNN-based model takes in unstructured input, it does not rely much on the expertise knowledge and pre-processing of the system. It can be readily applied to complex molecular systems. Due to these merits, GNN-based models are likely suitable for general-purpose applications like force-field development. However, there are some caveats in deploying the existing GNN-based models: (1) Many GNN-based models suffer issues like non-smooth cutoff when dealing with condensed-phase systems (particularly, with periodic boundary conditions) and long-range interactions; (2) Existing GNN-based models exclusively build upon pair-wise distances, thus may lack the expressivity for intrinsic higher-order interactions or properties (like chirality) compared to descriptor-based models; (3) GNN-based models are generally intensively over-parameterized, and the computational complexity is generally $\mathcal{O}(N^2)$, consequently, both the training and runtime of these models could entail prohibitive computational costs.

## 4. Model smoothness

No matter a descriptor-based or GNN-based model is exploited, the gradients produced by the model are not necessarily smooth enough, which may cause instabilities for numerical integration in simulations if the gradients are collected as forces in Newtonian mechanics. To attack the issue of irregular gradients while maximally harnessing the expressivity, one usually needs to impose certain regularization in order to achieve a "smooth" model. In the research of deep ANNs, Lipschitz continuity is a commonly adopted measure quantifying the smoothness of the model. A *K*-Lipschitz function is a continuous function which has gradients whose norm is no larger than a positive constant *K* everywhere. Given this definition, one is motivated to add a gradient-penalty term in order to restrict the model to be *K*-Lipschitz.[40] More recent studies revealed that the non-smoothness of ANNs can result from the fact that the weight matrices of ANNs become ill-conditioned during training.[41-42] Therefore, if one is able to properly "normalize" the weight matrices during training and prevent the ill-conditioning, the resulting model will be smooth. Spectral normalization[42] is such a technique working according to this principle, which normalizes the leading singular value of the weight matrices, and it has been shown that spectral normalization can significantly stabilize the training of generative adversarial networks (GANs)[43] and energy-based models (EBMs).[44] Besides, additional regularization techniques which could facilitate the training of the deep ANNs are often simultaneously implemented. Particularly, for over-parametrized molecular models, it is important to guarantee that the models are less prone to overfitting. For this purpose, one can directly borrow the general techniques developed in deep learning community, for instance, L2-regularization[45] and multi-task learning,[46] etc.

Another related issue is that ANNs usually work robustly only for inputs and outputs which are neither very small nor very large. However, in molecular modeling, the value of the output or label (e.g., energy or inter-particle distance) is often unbounded. To address this issue, one common approach is to rescale the output using suitable constants.[47] Under some settings, although the value of the output is unbounded, we may be only concerned with values within a certain range. If this is the case, one can first decide on a value range of interest, then discretize this range with a sufficient number of bins. A simplex over these discretized bins can then be predicted by the ANNs (e.g., via softmax).[34, 48]

In addition to the gradient regularizer, choice of activation functions is also critical to the performance and smoothness of the deep molecular models. Instead of the saturating functions like sigmoid or hyperbolic tangent, quasi-linear activation functions with stronger Lipschitz-continuity are preferred for molecular modeling. Such activation functions include but not limit to (shifted) softplus,[49-50] exponential linear units (ELU),[51] Swish,[52] etc, and there is growing evidence that these activation functions could help achieve desired performance on molecular property predictions.[35, 37]

## Section II. Deep supervised learning for molecular modeling and simulations

The aim of deep supervised learning is to find a model $f_\theta(\mathbf{x})$, which inputs $\mathbf{x}$ and predicts the corresponding label $A(\mathbf{x})$. The core of supervised learning lies in the acquisition of the labels. In machine learning, classification tasks and regression problems all belong to the realm of supervised learning. Supervised learning is perhaps the most widely exploited learning scheme for deep molecular modeling. For instance, there were several attempts to extract order parameters via supervised learning.[53-54] Besides, many studies tried to fit PES or free energy surfaces (FES) with regressive deep learning, and we will present several examples in the following.

### 1. Fitting potential energy surfaces

If the atomic positions $\mathbf{x}$ are chosen as the input, and the value of the corresponding potential energy $U(\mathbf{x})$ is provided as the label, one can formulate a regression problem where a model $f_\theta(\mathbf{x})$ is trained to minimize the loss function ( "loss function" in this paper is interchangeable with the "cost function" in machine learning literature) in Eq. (8),

$$L(\theta) = \langle |f_\theta(\mathbf{x}) - U(\mathbf{x})|^2 \rangle_\mathbf{x} \tag{8}$$

where $\langle A(\mathbf{x}) \rangle_\mathbf{x}$ stands for the expectation value of $A(\mathbf{x})$ over a sample set of $\mathbf{x}$. In this way, one is trying to fit a PES using a deep molecular model $f_\theta(\mathbf{x})$. At the first sight, fitting an energy function $U(\mathbf{x})$, the analytical form of which is known, by another unknown function $f_\theta(\mathbf{x})$ makes little sense. However, if the evaluation of $f_\theta(\mathbf{x})$ is significantly cheaper than $U(\mathbf{x})$, replacing $U(\mathbf{x})$ by $f_\theta(\mathbf{x})$ will help reduce a great deal of computational effort. This is often the case in quantum chemistry and *ab initio* molecular dynamics (AIMD),[55] where the evaluation of even a single-point energy entails heavy computation. Traditionally, one has to trade the accuracy for computational speed in order to apply AIMD or quantum-level calculations in complex molecular systems. While by virtual of deep molecular models, one can now perform simulations or single-point energy calculations on a surrogate deep PES, thus striking a good balance between accuracy and computational complexity.

Although fitting a PES appears to be a simple deep learning task, the model architectures and the training objectives have undergone dramatic evolution along history. One of the earliest attempts along this line is BPNN.[21] As introduced earlier, BPNN encodes the molecular configuration into atom-wise descriptors. Based on these AEV descriptors, a set of MLPs output the atomic energy for each atom, the summation of which gives the overall potential energy. BPNN is trained mainly through energy matching, i.e., Eq. (8). Due to the relative demanding requirements of suitable AEV descriptors, applications of BPNN and its following variants were mostly limited to simple systems, for example, small organic molecules[22] or homogeneous material systems.[27]

As another descriptor-based model, DPMD[2] adopts a different feature-engineering strategy so that it is capable of handling larger and even more complex systems. Besides, DPMD included and emphasized more on a force-matching term in the training objective in addition to the energy-matching term. Apparently, the force-matching term can be viewed as one kind of multi-task learning, thus helps regularize the gradients or smoothness of the model. We remark here that the force-matching brings additional benefit from the perspective of data augmentation. As introduced, the foundation of a successful supervised learning is the acquisition of labels, so data augmentation is a very important technique for supervised learning. Given a fixed dataset of $\{\mathbf{x}, U(\mathbf{x}), \nabla_\mathbf{x} U(\mathbf{x})\}$, predicting the force at $\mathbf{x}$ is roughly equivalent to predicting the energy at an adjacent point, that is, $U(\mathbf{x} + d\mathbf{x})$. Therefore, inclusion of the force-matching term implicitly augments the training data. Other techniques for explicit data augmentation, such as active learning and adaptive sampling, were also introduced by related studies.[37]

On the other end, GNN-based models have seen rapid development during the last couple of years. For example, SchNet[35], DTNN,[36] PhysNet[37] and DimeNet[56] raised the bar of the performance of deep molecular models on many benchmark tasks. Besides, the authors of PhysNet found it necessary to incorporate known physics laws in order to achieve correct asymptotic behavior in the long range. Specifically, they additionally predicted the atomic partial charges as part of the overall training objective. From the viewpoint of physics, predicting atomic charges is equivalent to giving the electrons a separate description, and also allows the incorporation of the Coulomb's law to capture the correct long-range interactions.[37] Similar ideas were also employed by TensorMol[23] which separately predicted van der Waals interactions which help correct the short-range asymptotic behavior. We comment here that such change of the training objective makes sense in that asymptotic behavior can be regarded as a special type of inductive biases, and designing loss functions that explicitly account for the asymptotic behavior will help lead to better models.

Reflecting the history from BPNN to PhysNet, in addition to designing ever more complex molecular models, we should re-think how to *correctly* exploit them. Of course we can regard the deep ANN as a sheer black-box function approximator, and use it to directly fit the overall energy function (as in BPNN). Indeed, the training objectives introduced in the above examples are all empirical risks defined by the users. For supervised learning, the choice of the empirical risk is often vital to the performance of the final model. A recent study[57] indicates that if merely based on the empirical risks like Eq. (8), the partition of the overall energy into atomic contributions could be arbitrary and non-interpretable as evidence of overfitting for models with sufficient capacity. Therefore, before deciding on an empirical risk as the loss, one should reflect whether there is any alternative learning objective that can lead to better sample efficiency and model transferability.

## 2. Fitting potential of mean force

To investigate molecular systems, one is often not interested in the potential energy $U(\mathbf{x})$, but the free energy or potential of mean force (PMF), $F(\mathbf{q})$, of some reduced descriptors $\mathbf{q}$ like coarse-grained (CG) variables [58]:

$$p(\mathbf{q}) = \frac{\int e^{-\beta U(\mathbf{x})}\delta(\mathbf{q} - \mathbf{q}(\mathbf{x}))d\mathbf{x}}{\int e^{-\beta U(\mathbf{x})}d\mathbf{x}} = \frac{e^{-\beta F(\mathbf{q})}}{Z} \quad (9)$$

$$F(\mathbf{q}) = -\frac{1}{\beta}[\log p(\mathbf{q}) + \log Z] \quad (10)$$

where $Z = \int e^{-\beta U(\mathbf{x})}d\mathbf{x}$ is the partition function, $\beta$ is the inverse temperature and $\delta$ the Dirac delta function. If $\mathbf{q}$ corresponds to the slowly changing variables governing the process of interest, $F(\mathbf{q})$ becomes a CG description of the original thermodynamic system, and simulations performed under $F(\mathbf{q})$ are generally much faster than those run on the fine-grained (FG) potential $U(\mathbf{x})$. Therefore, Equations (9-10) are also known as the principle of thermodynamic consistency for coarse graining.[59]

However, it is often challenging to derive a reliable analytical form for $F(\mathbf{q})$ given access to samples drawn from $U(\mathbf{x})$, particularly when $\text{Dim}(\mathbf{q})$ is large ($\text{Dim}(\cdot)$ is short for the dimensionality). This problem is particularly pronounced in the realm of coarse graining. Considered that ANNs may offer extra flexibility and expressivity to this end, several recent studies proposed supervised learning approaches to fit $F(\mathbf{q})$ by ANNs.[60-62] For instance, Schneider *et al.* gridded the possibly high-dimensional CG variable space, and adopted enhanced sampling techniques to generate samples over the $M$ grids. Based on these samples denoted by $\{\mathbf{q}^{(i)}\}_{i=1,\ldots,M}$, the corresponding free energy, $F(\mathbf{q}^{(i)})$, and the mean force, $-\nabla_{\mathbf{q}^{(i)}}F(\mathbf{q}^{(i)})$, at these gridded points can be estimated, and their values were used to train an ANN $F_\theta(\mathbf{q})$ in order to fit the FES. The training objective is to minimize the empirical risk in Eq. (11),

$$L(\theta) = \langle |F_\theta(\mathbf{q}^{(i)}) - F(\mathbf{q}^{(i)})|^2 \rangle_i \tag{11}$$

Wang *et al.* further extended this idea into particle-based coarse graining models which are widely adopted for bio-molecular systems.[4] Since the FES of a particle-based model should satisfy the inductive biases of molecular models, Wang *et al.* trained an CG potential $F_\theta(\mathbf{s})$ using one kind of deep molecular model (called "CGnet") to represent the FES of the CG particles, and trained the CGnet using the mean-force information,

$$L(\theta) = \langle \left\| \nabla_{\mathbf{q}^{(i)}} F_\theta(\mathbf{q}^{(i)}) - \nabla_{\mathbf{q}^{(i)}} F(\mathbf{q}^{(i)}) \right\|_F^2 \rangle_i \tag{12}$$

The authors of CGNet found that imposing physics-based restraints into the overall model is essential for a reasonable performance. Inclusion of physics-based restraints in $F_\theta$ is straightforward thanks to the additive compositionality of the energy function. Noteworthy, both Eq. (11) and Eq. (12) coined a regression problem of $F(\mathbf{q})$ via empirical risk minimization (ERM). If $\mathbf{q}^{(i)}$ is drawn from the equilibrium distribution, Eq. (12) is also known as score-matching and equivalent to minimizing the Fisher's divergence.[63] However, since the training data is manually curated and generally off-equilibrium, ERM is very sensitive to the sample coverage and the formula of the loss function. For example, replacing the L2-norm or the Frobenius norm in Eqs. (11-12) with other norms, or adding weights to the norm calculation would dramatically influence the model performance, thus the choice of proper dataset and loss function can be very tricky. Besides, fitting $F(\mathbf{q})$ in a regression fashion has several potential drawbacks: It necessitates gridding the space of $\mathbf{q}$, which would be computationally prohibitive for large $\text{Dim}(\mathbf{q})$; Beside, it is rather data-inefficient because calculating $F(\mathbf{q})$ at one point requires a large amount of samples from $U(\mathbf{x})$ at the neighborhood of $\mathbf{q}$, and requires calculation of the mean forces which is usually not done in practice. We will show in the following section that, instead of ERM, distribution learning and density estimation methods are more mathematically proper ways to deal with these problems (i.e., coarse graining or inferring FES), and they are all central topics of unsupervised learning.

**Section III. Deep unsupervised learning for molecular modeling and simulations**

Although most successful applications of deep molecular models are centered at supervised learning, many challenging problems in molecular modeling and simulations are of unsupervised nature. For instance, dimensionality reduction and clustering are two commonly encountered unsupervised learning problems in molecular simulations. Besides, density estimation is another important unsupervised learning problem which is widely involved in molecular modeling scenarios like free energy calculation and coarse graining. To tackle these problems via deep learning, one needs to translate the physical or mathematical nature of the problems into optimizable (or possibly, variational) objectives. In the following we will present several examples in which deep unsupervised learning was implemented to eliminate possible subjective biases and offer more flexible solutions to these problems.

**1. Dimensionality reduction and clustering**

Many molecular dynamic processes in chemistry and biology, e.g. chemical reaction, protein folding and ligand binding etc., can be described by a clustered free energy landscape (FEL)[64] consisting of many local free energy minima which correspond to the metastable states. This picture of a clustered FEL (or the metastability) is the cornerstone of many kinetic models dealing with diffusive and complex dynamics, and a simplified and informative visualization of complex FEL amenable for downstream tasks such as clustering is often required by these kinetic models. Zhang *et al.* recently proposed a parametric learning approach, called Information Distilling of Metastability (IDM), to perform clustering in the meantime of reducing the dimensionality for molecular systems.[8] IDM is end-to-end differentiable thus scalable to ultra-large dataset. Remarkably, IDM neither requires a cherry-picked distance metric nor the ground-true number of clusters defined *a priori*, and it can be used to unroll and zoom-in the hierarchical FEL with respect to different timescales.

IDM is based on the idea of information co-distillation originated from machine learning community.[65-66] Given two observations $\mathbf{x}$ and $\mathbf{x}'$ belonging to the same metastable state, one can find a function $\chi$ capturing what is in common between them by maximizing the mutual information (MI), $I(\chi(\mathbf{x}), \chi(\mathbf{x}'))$. In order to avoid the trivial solution of $\chi$ being the identity function, Zhang *et al.* confines $\chi$ to the family of classification functions (or indicator functions) with finite categories. Consequently, the entropy of $\chi$ is upper bounded, hence the information is bottlenecked and distilled. Clustering can thus be achieved by maximizing the following closed-form mutual information:

$$I_\beta(\chi(\mathbf{x}), \chi(\mathbf{x}')) = \sum_{c,c'=1}^{K} \langle \chi_c(\mathbf{x}), \chi_{c'}(\mathbf{x}') \rangle \ln \frac{\langle \chi_c(\mathbf{x}), \chi_{c'}(\mathbf{x}') \rangle}{\langle \chi_c(\mathbf{x}) \rangle^\beta \langle \chi_{c'}(\mathbf{x}') \rangle^\beta} \qquad (13)$$

where a controlling hyper-parameter $\beta$ is introduced without loss of generality. $\chi_c(\mathbf{x})$ denotes the $c$-th entry of $\chi$ corresponding to the probability of $\mathbf{x}$ belonging to cluster $c$. Intuitively, Eq. (13) means that the clustering mapping of $\mathbf{x}$, namely $\chi(\mathbf{x})$, should be maximally informative with its temporal neighbors. In practice, this is obtained when $\chi(\mathbf{x}) \approx \chi(\mathbf{x}')$ for arbitrary $\mathbf{x}$.

Since Eq. (13) is a variational objective, one is motivated to adopt ANNs as the clustering function $\chi$ because they usually offer sufficient variational flexibility over high dimensional data. In order to "hit two birds with one stone", that is, reduce the dimensionality while clustering, IDM employs the Matching Networks,[67] a special deep neural network architecture consisting of two twin networks and mapping two different vector spaces onto a same vector space. One network projects the molecular configuration $\mathbf{x}$ to a low dimensional *matching space* which serves as the reduced representation of FEL. The other twin-network maps constant vector **c**, which acts as the preimage of a cluster centroid, to the same matching space (Fig. 2A). The final soft clustering labels $\chi(\mathbf{x})$ are then defined by the similarity between the projected images of **x** and **c** on the matching space. Noteworthy, IDM does not require a reference to the exact number of ground-true clusters. Instead, IDM robustly clusters the data as long as the maximal allowed number of clusters, $K$ in Eq. (13), is large enough. Besides, IDM yields soft cluster labels $\chi(\mathbf{x})$, which are preferred in many scenarios to hard ones. Finally, trained upon $N$ decreasing temporal neighborhood sizes ($\tau_1 \gg \tau_2 \gg \cdots \gg \tau_N$), IDM allows one to zoom-in the FEL with increasing time resolutions, which is equivalent to performing a top-down divisive clustering, hence IDM is able to extract the hierarchy of the FEL according to timescales rather than geometry-based metrics.

As an example, IDM was divisively performed on Ala2 for illustration (Figs. 2B and 2C).[8] Three timescales were chosen for training: (1) $\tau = 200$ ps, (2) $\tau = 20$ ps, and (3) $\tau = 2$ ps. In all the three cases, the maximal allowed cluster number $K$ was set to be 16. On the longest timescale ($\tau = 200$ ps), IDM partitions all conformations into two metastable states (Fig. 2B, panel 1), corresponding to the *cis/trans*-isomers of the torsional angle $\phi$. This is in good agreement with kinetic modeling result that isomerization of $\phi$ is much slower than $\psi$. As expected, the reduced representation obtained by IDM at this timescale only preserves two distinguishable metastable states (Fig. 2C, panel 1). When progressing to a smaller timescale of 20 ps, IDM further divides the conformations into four clusters, as reported in previous sections (panel 2 in Fig. 2B), and the reconstructed FEL now consists of 4 distinguishable metastable states (panel 2 in Fig. 2C). At an even smaller timescale, $\tau = 2$ ps, IDM categorized all the conformers of Ala2 into 5 metastable states (Fig. 2B, panel 3) and learns a reduced representation accordingly (Fig. 2B, panel 3). This example shows that IDM enables one to zoom-in the FEL with increasing time resolutions and track the hierarchy of metastable states accordingly.

## 2. Density estimation

In the previous section, we discussed the pros and cons for constructing a FES $F(\mathbf{q})$ given access to samples drawn from $U(\mathbf{x})$ via supervised empirical risk minimization. An alternative (perhaps also more reasonable) view of inferring $p(\mathbf{s})$ is provided by statistical modeling and unsupervised learning, in which density estimation of high-dimensional data has been a long-standing goal.[19, 68] Conventionally, if **q** is low-dimensional (say, $\text{Dim}(\mathbf{q}) \leq 3$), non-parametric methods like kernel density estimation (KDE)[69] can be adopted to infer $F(\mathbf{q})$, but they become quickly infeasible as $\text{Dim}(\mathbf{q})$ increases. Recently, a burst of work sparks new ideas to exploit deep learning to this end, giving rise to an active research field of generative learning and models including variational auto-encoders,[70] GANs,[43] auto-regressive,[71] and normalizing-flow models.[72] Nevertheless, these methods are mostly footed on certain simple prior distributions, hence they are likely to suffer from issues like mode-dropping and often assign probability mass to areas unwarranted by the data.[73] Worse still, these generative models generally do not comply with the inductive biases of molecular systems. In contrast, generative learning with EBMs has an even longer history,[74-76] which in principle can fit arbitrarily complex distributions due to the flexibility and plasticity of the energy landscapes.[44, 77] Based on EBMs, Zhang *et al.* recently proposed a variational approach, called variational inference of free-energy (VIFE),[3] to approximate the FES ($F(\mathbf{q})$) by parametric models and without supervision.

Specifically, let us denote the approximate free energy function and the associated probability distribution as $F_\theta(\mathbf{q})$ and $p_\theta(\mathbf{q})$ respectively (where $\theta$ are optimizable parameters), and define the Kullback-Leibler divergence, $D_{\text{KL}}(p||p_\theta)$, between $p$ and $p_\theta$. $D_{\text{KL}}(p||p_\theta)$ is a strict divergence in that $D_{\text{KL}}(p||p_\theta) \geq 0$ where the equality holds if and only if $p = p_\theta$, hence can be used as a variational objective.[78-80] Particularly, the gradient of $D_{\text{KL}}(p||p_\theta)$ w.r.t. $\theta$ takes the following form in Eq. (14),

$$\nabla_\theta D_{\text{KL}}(p||p_\theta) = \langle \beta \nabla_\theta F_\theta(\mathbf{q}) \rangle_{p(\mathbf{q})} - \langle \beta \nabla_\theta F_\theta(\mathbf{q}) \rangle_{p_\theta(\mathbf{q})} \qquad (14)$$

If one has access to, say, equilibrium MD samples drawn from $U(\mathbf{x})$, the distribution of which is denoted as $p_{\text{FG}}(\mathbf{q})$, one can use $p_{\text{FG}}$ to approximate $p$ and optimize $F_\theta$ w.r.t. a surrogate objective $D_{\text{KL}}(p_{\text{FG}}||p_\theta)$. Since the analytical forms of $F_\theta(\mathbf{q})$ and $\nabla_\mathbf{q} F_\theta(\mathbf{q})$ are both available, one can perform CG simulations on $F_\theta(\mathbf{q})$ to estimate $\langle \nabla_\theta F_\theta(\mathbf{q}) \rangle_{p_\theta(\mathbf{q})}$. Due to the simplicity of $F_\theta(\mathbf{q})$, the CG simulations are usually computationally economic and converge relatively fast, hence brute-force simulations via Monte Carlo (MC) sampling or Langevin dynamics generally suffice.

As an example, VIFE was benchmarked on a 2-dimensional 3-well toy model, the PES of which is shown in Fig. 3A.[3] Langevin dynamics simulation was first performed on $U(x, y)$ and yielded 50,000 samples which were used as $p_{\text{FG}}$. The CG variable $\mathbf{q}$ was identical to $(x, y)$, and $F_\theta$ was built in order to reproduce $U(x, y)$ rather than coarse graining. An ANN representing $F_\theta$ was trained according to Eq. (14). It can be seen that $D_{\text{KL}}$ between the two distributions quickly diminished during training, and the optimization of $F_\theta$ converged within 50 training epochs (each epoch is composed of 100 mini-batch optimization steps). Compared to the original PES, the optimized $F_\theta$ preserves all the three local minima correctly, showing no signs of mode-dropping (Fig. 3B). This is an advantage of EBMs over many other generative methods because metastable states usually play functionally important roles for bio-molecules, and dropping any of them during may lead to unreasonable models. Besides, the overall contour and landscape of $F_\theta$ also resemble $U(x, y)$, especially on regions with higher densities. Now that the analytical form of $F_\theta$ was obtained based merely on samples from $U(x, y)$ (rather than knowing the mathematical form of $U(x, y)$), one can locate the free energy minima and cluster samples. The simulation samples were minimized over $F_\theta$, and all the samples finally fell into three distinct local minima as is shown by different colored symbols in Fig. 3C, and the noisily distributed samples were indeed assigned to different metastable states quite reasonably.

In addition to the above example, density estimation and generative learning via EBMs can also be adopted for prediction tasks. For example, Ingraham *et al.* proposed a generative model to predict folded protein structures.[81] They first constructed a Markov Random Field (MRF) via deep ANNs using coevolution information such as homologous sequences to quantify the effective energy of a specific protein conformation. Meanwhile, they also parametrized a Langevin dynamics with a trainable diffusion field, in combination with the drifting field governed by the MRF. Running this Langevin dynamics thus allows sampling of protein conformations given a particular amino acid sequence. Because the overdamped Langevin dynamics simulator (both the drifting field and diffusion field) is end-to-end differentiable, the authors termed it as a "Differentiable Simulator", and they trained it by maximizing the likelihood of the generated conformations w.r.t. the known folded structures. Similarly, AlphaFold[48] also predicts a PMF for the protein conformations based on deep learning over coevolution information. AlphaFold takes advantage of the compositionality of EBMs, and decomposed the PMF into additive and interpretable components including, e.g., one term accounting for the pair-wise distances between residues and one for the backbone torsional angles. In the Differentiable Simulator and AlphaFold, how to perform sampling within EBMs is worthy of particular discussion. In the Differentiable Simulator, the authors trained a sampling engine in addition to the energy function. In contrast, instead of performing equilibrium sampling (as by Langevin dynamics), AlphaFold predicts the protein structures by direct gradient descent over the energy function thus avoiding the explicit sampling procedure, but such generating procedure no longer allows maximum likelihood optimization as in the Differentiable Simulator. Indeed, many recent studies were trying to develop gradient-descent-like inference method which also supports maximum likelihood optimization of EBMs for prediction tasks.[82-83] Nonetheless, sampling is still a critical obstacle for most practitioners of EBMs. Once the sampling issue is better solved, energy-based generative learning would find more applications in other challenging molecular modeling tasks such as inverse coarse graining. In the next section, we will discuss how reinforcement learning could offer advanced solutions to the challenging sampling problem.

## Section IV. Deep reinforcement learning for molecular modeling and simulations

Under settings of a conventional reinforcement learning (RL) task, the agent is trained to find the optimal strategy, $\pi(a,s)$ which dictates the probability of choosing an action $a$ at the state $s$, in order to maximize a specific (task-related) reward function $R(s)$.[84] Usually the RL problem consists of a sequence of decision making, leading to a trajectory of state-action pairs along with the intermediate and final accumulated rewards. Due to its definition, RL is arguably the most general framework for machine learning. It has drawn increasing attention from deep learning community and seen the most burgeoning development in recent years. Although RL seems not directly related to molecular modeling and simulations in light of its definition, the philosophy and mathematical framework of RL are deeply connected to statistical physics.

One of the mathematical cornerstones of RL is *expectation maximization*,[85-86] which arises from the fact that one can only manage to maximize the *expected* reward due to the stochastic and noisy environment. In order to calculate the expectation, sampling according to the policy is thus inevitable. Consequently, the sampling issue suffered by molecular simulations is also pronounced in RL. Moreover, since the policy is ever-changing during training, the dataset is always dynamic in RL. Therefore, how to handle partial sampling and dynamic dataset remains as central research topics in RL, which may also help address similar issues encountered in molecular modeling and simulations.

Besides, the concept of "reward" is quite ambiguous. Actually the reward function $R(s)$ can be defined arbitrarily. Therefore, if one can define a proper reward specific to the concerned problem, one can naturally deploy methods and techniques from RL to solve their own problems. For instance, some adaptive sampling strategies were newly proposed in the spirit of RL.[61, 87] Although the authors did not rigorously formulate their methodologies in terms of RL, they all signified that their methods take a flavor of RL. In fact, these adaptive sampling methods all belong to a category called curiosity-based RL,[88] which is an active research field in machine learning community.

Perhaps more interesting connections between RL and EBMs can be revealed by the recently developed deep RL algorithms for continuous state and action spaces.[89-90] Since the policy is defined over a continuous space of actions, it is natural to parametrize the policy in the form of energy functions, so that the probability of picking a certain action is governed by the Boltzmann distribution. In turn, we can also regard the Boltzmann distribution induced by the energy function in EBMs as a special "policy", under which the sampler (an imaginary agent) is traveling through the phase space. In the following we will present several examples which incorporate these interesting ideas and techniques from RL in molecular modeling and simulations.

### 1. Actor-Critic enhanced sampling

Many dynamic events of interest in molecular systems, e.g. phase transitions, protein folding and ligand binding, are embedded in a complex free energy landscape where long-lived metastable states are separated by kinetic bottlenecks (i.e. free energy barriers) which impede the transitions.[91] Therefore, timescales of these rare events are well out of the reach of brute-force simulations, and enhanced sampling methods are largely needed to this end. Zhang *et al.* recently re-formulated the enhanced sampling problem as a distribution learning problem, and proposed a *minimax* game as in GANs to solve it. The proposed approach is called targeted adversarial learning optimized sampling (TALOS).[6]

TALOS simultaneously trains a bias potential, $V_\theta(\mathbf{x})$, which is applied to a simulation engine like MD (which plays a role as the "generator" in GANs), and a discriminator $D_w(\mathbf{q}(\mathbf{x}))$ that attempts to differentiate samples generated by the biased sampler from those drawn from the desired target distribution $p(\mathbf{q})$ (where $\mathbf{q}$ is a descriptor as a function of $\mathbf{x}$). Note that $\theta$ and $w$ denote the trainable parameters of the deep molecular models. The biased sampler receives a reward when it fools the discriminator, whereas the discriminator is rewarded when it successfully classifies a sample as from the "real" target distribution or the "fake" biased simulation. The training ends when the *Nash equilibrium* of the game is reached. Inspired by Wasserstein-GAN [40], the entire training process of TALOS is equivalent to solving the following *minimax* problem:

$$\min_{V_\theta} \max_{D_w \sim \mathbb{L}1} \left[ \langle D_w(\mathbf{q}) \rangle_{p(\mathbf{q})} - \langle D_w(\mathbf{q}) \rangle_{p_\theta(\mathbf{q})} \right] \quad (15)$$

where $\max_{D_w \sim \mathbb{L}1} \left[ \langle D_w(\mathbf{q}) \rangle_{p(\mathbf{q})} - \langle D_w(\mathbf{q}) \rangle_{p_\theta(\mathbf{q})} \right]$ is the Wasserstein-1 distance[92] between the target distribution $p$ and the biased distribution $p_\theta \propto \exp\left(-\beta(U(\mathbf{x}) + V_\theta(\mathbf{x}))\right)$. $D_w(\mathbf{q})$ belongs to the family of 1-Lipschitz ($\mathbb{L}1$) functions mapping the descriptor space $\mathbf{q}$ to the real space. However, the analytical form of the optimal $D_w$ is unknown, so a neural network conforming to a Lipschitz restraint, parametrized by $w$, is employed to approximate it following the gradient:

$$\nabla_w L(w) = \langle \nabla_w D_w(\mathbf{q}) \rangle_{p_\theta} - \langle \nabla_w D_w(\mathbf{q}) \rangle_p \quad (16)$$

Once the discriminator $D_w$ is fixed, the bias potential $V_\theta(\mathbf{x})$ is optimized following the gradient:

$$\nabla_\theta L(\theta) = \langle \left(D_w(\mathbf{q}) - \langle D_w(\mathbf{q}) \rangle_{p_\theta}\right) \beta \nabla_\theta V_\theta(\mathbf{x}) \rangle_{p_\theta} \quad (17)$$

where $\beta$ is the inverse temperature factor. Equations (16-17) bear intriguing similarity with the Actor-Critic algorithm,[93] hence rendering TALOS an interpretation in terms of RL: Equation (16) is parallel to the training of a Critic which is used to approximate the value function of being in a state $\mathbf{q}(\mathbf{x})$; whereas Eq. (17) is analogous to the famous REINFORCE algorithm (also known as the policy gradient or Expectation Maximization) [85] which maximizes the expected value $\langle D_w(\mathbf{q}) \rangle_{p_\theta}$ for the Actor who visits state $\mathbf{q}(\mathbf{x})$ with a probability proportional to $\exp\left(-\beta(U(\mathbf{x}) + V_\theta(\mathbf{x}))\right)$. With this insight, one can develop variants of TALOS by defining other task-specific value functions and manipulate the PES accordingly.

Noteworthy, TALOS does not require the bias potential and the target distribution to operate in the same space, providing additional flexibility in designing and applications of the method. As an example, TALOS was applied to an all-atom simulation of the S$_N$2 reaction $Cl^- + CH_3Cl \leftrightarrow CH_3Cl + Cl^-$ (Fig. 4A).[6] The difference between the two C-Cl bond distances $q^* = d_1 - d_2$ serves as a 1-dimensional descriptor (Figs. 4A and 4B) to define the target distribution (Fig. 4C), whereas a 2-dimensional bias potential as a function of $(d_1, d_2)$ is constructed. During training, the bias potential was updated in an interpretable manner similar to metadynamics: The critic appears to be able to "feel" the difference between the target and current distributions so that a positive bias potential was added to penalize where was oversampled and a negative one to encourage where was undersampled (Figs. 4D and 4E). As a consequence, the system was driven to the predefined target distribution giving rise to faster chemical transitions (Fig. 4D). Intriguingly, although the target descriptor is 1-dimensional, the bias potential optimized by TALOS is complementary to the 2-dimensional FES (Fig. 4E). Given that the bias potential and target distribution can be flexibly chosen, TALOS is applicable to a variety of situations where inference or modification to the Hamiltonian is needed in order to achieve a specific distribution in addition to enhanced sampling.

## 2. Reinforced multi-scale molecular modeling

As introduced in the previous section, VIFE can be used to perform density estimation or coarse graining given equilibrium samples from a fine-grained (FG) simulation, denoted by $p_{FG}$. However, in a more common setting where equilibrium FG samples are not available beforehand, one has to perform sampling over $U(\mathbf{x})$ from scratch. Since $\text{Dim}(\mathbf{x})$ is usually very large, estimating ensemble averages over $p(\mathbf{q})$ in Eq. (14) is often infeasible for brute-force FG (e.g., all-atom or *ab initio*) simulations.

Zhang *et al.* showed that VIFE combined with RL, provides a new solution to the this sampling problem, and called this new method reinforced VIFE (RE-VIFE).[3] Inspired by the off-policy RL,[90] RE-VIFE exploits a two-timescale learning scheme, where a bias potential $V_\phi(\mathbf{q})$ with $\phi$ denoting optimizable parameters (equivalent to a policy function in RL) is separately trained in addition to $F_\theta(\mathbf{q})$ which can now be viewed as a value function in the spirit of Actor-Critic RL.[94] In RE-VIFE, a target distribution $p_T(\mathbf{q})$ is defined according to the value function $F_\theta(\mathbf{q})$, and the target distribution encourages the sampler to explore less-visited regions. To move the sampling distribution towards $p_T$, the bias potential $V_\phi(\mathbf{q})$ can be optimized by minimizing a strict divergence, for instance, $D_{\text{KL}}(p_T||p_\phi)$,

$$\nabla_\phi D_{\text{KL}}(p_T||p_\phi) = \langle \beta \nabla_\phi V_\phi(\mathbf{q}) \rangle_{p_T} - \langle \beta \nabla_\phi V_\phi(\mathbf{q}) \rangle_{p_\phi} \tag{18}$$

where $p_\phi$ denotes the Boltzmann distribution under $(U + V_\phi)$. For example, if one chooses a well-tempered target distribution (Eq. (19)),

$$p_T = p_\theta^{\frac{1}{\gamma}} \propto \exp\left(-\frac{\beta}{\gamma} F_\theta\right) \tag{19}$$

where $\gamma > 1$ is the well-tempering factor,[95] then the optimal $V_\phi$ will converge to the well-tempered free energy. In terms of imitation learning,[96] $F_\theta(\mathbf{q})$ plays the role of a leader that coins a moving target based on the current density estimation, while $V_\phi(\mathbf{s})$ learns to tune the policy in order to catch up (Fig. 5A). A dramatic advantage of such leader-chaser scheme lies in the fact that $F_\theta$ along with $p_T$ is constructed based on the samples drawn from simulations under $V_\phi$, so $p_T$ and $p_\phi$ always share substantial overlap; otherwise $D_{\text{KL}}$ would fall victim to the notorious vanishing gradient issue.[40]

As an example, consider the problem of building a CG model for a prototypical bio-molecular system, alanine dipeptide (Ala2) in explicit water, where the backbone torsional angles, $\mathbf{q} = (\phi, \varphi)$, were chosen as the CG variables (Fig. 5B).[3] Assuming that no FG samples were provided *a priori*, one has to perform sampling over the all-atom model from scratch. However, since isomerization of $(\phi, \varphi)$ is a rare event due to relatively high free energy barriers, brute-force FG simulations of Ala2 converge too slowly to obtain an accurate estimate of $\langle \nabla_\theta F_\theta(\mathbf{q}) \rangle_{p_{\text{FG}}}$. RE-VIFE was thus adopted to enhance the FG sampling over $\mathbf{q}$, and two simulations were launched simultaneously: one FG (all-atom) MD simulation under a bias potential $V_\phi(\mathbf{q})$, and a CG MC simulation over $F_\theta(\mathbf{q})$. Figure 5C showcases a 1-ns FG simulation trajectory for torsions $\phi$ and $\varphi$ produced by vanilla MD was presented in contrast to those produced by simulations biased by $V_\phi$. It can be found that the brute-force MD can hardly produce equilibrium samples covering all important metastable states in this short simulation length. In contrast, RE-VIFE helps enhance the sampling of FG models in that the isomerization of torsion $\phi$ becomes reasonably frequent. Such a boosted sampling efficiency can be explained by the optimized bias potential $V_\phi$ via RE-VIFE (Fig. 5D), which appears complementary to the ground-true FES of $\mathbf{q} = (\phi, \varphi)$ thus effectively flattens the FES. On the other hand, as samples from $V_\phi$ are better representatives of $p_{\text{FG}}$ (excellently covering both the free-energy minima and the transition regions), they can be reliably used to optimize the CG model. The final CG model ($F_\theta$) optimized via RE-VIFE for Ala2 in explicit water is shown in Fig. 4E. Noteworthy, $F_\theta$ captures all known metastable states of Ala2 w.r.t. $(\phi, \varphi)$ (i.e., no mode-dropping), and quantitatively agrees well with the ground truth (Fig. 5E), demonstrating that simulations on multiple scales can be bridged by RE-VIFE and that CG models can be reliably inferred even without access to FG samples *a priori*.

## Outlook

Despite the encouraging progress so far, there are still many obstacles limiting the development and applications of deep molecular models. As a closing to this review, we will enumerate some issues that cause wide concerns and critics over the role of deep learning in molecular modeling and simulations, and outlook possible solutions to address these issues.

**1. Transferability**

Although there are already attempts using deep molecular models to construct deep PES, unfortunately, they have proved limited transferability or "generalizability" in practice. In other words, to establish a deep PES for a new system, one has to build up the model from scratch, including choosing a proper model architecture, creating and curating the training data, optimizing and validating the model etc. Frustratingly, it remains hard to directly re-use models previously trained by others. Such non-transferability is very unfriendly to users, thus severely limits the usage of current deep PES models. Though the model architectures have evolved to be more transferable than ever before, the transferability of a deep PES is still hindered by the empirical risk which is used as the learning objective.

One may argue that the PES can be fitted in the same fashion according to Eq. (8) for different systems, so the learning objective as in Eq. (8) is transferable. But since we have only limited amount of training samples, a model trained according to an improper objective is likely to fall victim to overfitting due to low sample efficiency. Overfitting makes the model performing well on a specific system to be possibly completely wrong for another. As explained earlier, directly fitting of the overall energy may not be a good empirical risk in terms of transferability. An objective suitable for transferable learning should encourage the model to discover common patterns or rules across different systems and make predictions mainly on them.

Indeed, incorporation of some recurring patterns and rules has led to the success of classical force fields.[97] In the force fields, interactions are described by separate but fixed functional forms. Besides, force fields usually include terms responsible for the correct asymptotic behavior. For example, Coulomb interactions are generally present in order to describe the long-range interactions, whereas van der Waals-like terms are included to account for the Pauli exclusion. Therefore, force fields successfully capture the common patterns and rules for various molecular systems, thus exhibiting flexible transferability. Besides, the number of parameters of force fields are usually limited, and such models are not prone to overfitting. Inspired by these observations, one may introduce deep molecular models to extend the scope of known force fields. As an example, PhysNet predicts the atomic partial charges based on nuclei positions during training, which can be regarded as "re-parametrizing" the static atomic partial charges with floating ones. Instead of directly predicting the overall energy, the re-parametrized models can alternatively predict the correct parameters for specific interaction functions (like the charges in Coulomb's law), thus are possibly more transferable and sample-efficient. We think that re-parametrization and development of force-field-like deep PES models (like PhysNet and TensorMol) might be a promising direction for future research.

**2. Learn more, with less**

The sample efficiency is another critical issue faced by existing deep molecular models, particularly in the realm of supervised learning like fitting deep PES. Usually fitting a deep PES for a condensed-phase molecular system requires a huge amount of training data, that is, samples of the single-point energy from a high-dimensional configuration space. Since computing the single-point energy is generally expensive, one thus desires to train a reliable deep PES model with as few samples as possible. However, it is non-trivial for current deep molecular models to achieve this goal due to the following reasons:

(1) The deep molecular models are usually heavily over-parametrized. Mathematically, the parameter space of a deep ANN is usually too vast, and the capacity of the resulting model is excessive.[45] As a result, to find an optimal model with good generalization ability for a specific regression task requires a lot of instances.

(2) The current deep molecular models are not readily transferable. As explained above, one cannot straightforwardly make use of existing or pre-trained models due to the non-transferability issue. Every time one hopes to fit a new deep PES, one needs to perform everything from scratch, and cannot exploit others' previous efforts and results.

In deep learning community, there are two useful approaches to increase the sample efficiency for the models. The first one is pre-training plus fine-tuning, also known as transfer learning.[98] Specifically, one can pre-train the model over several related tasks, thus obtaining a model with near-optimal values for most parameters. Then for a specific new task, a satisfying model can be obtained by merely fine-tuning a small number of parameters (usually corresponding to the last several layers next to the output layer). Mathematically speaking, transfer learning effectively reduces the number of parameters which needs optimization for a specific task, thus increases the sample efficiency of the model meanwhile preserving the overall expressivity. This strategy has been shown extremely successful in computer vision[99] and natural language modeling.[100] Clearly, if one wishes to implement transfer learning for molecular modeling, one has to select transferable models and suitable transferable learning objectives which can be shared by various molecular systems.

The other promising approach is meta-learning, also known as learning to learn.[101-103] Similar to transfer learning, in meta-learning, the model is also first trained upon a number of related tasks. Different from transfer learning which optimizes the model with a normal supervised learning objective, meta-learning generally optimizes the model w.r.t. its ability of generalization.[67, 104-105] In other words, meta-learning aims to train a model so that it can fast adapt to new samples to fulfil a new task within only a few optimization steps. Transfer learning can be viewed as a special type of meta-learning. Compared to transfer learning, meta-learning is particularly useful in one- or few-shot learning, which means that the model can make reasonable predictions in a new task given only one or a few instances. Both transfer learning and meta-learning could be promising in improving the sample efficiency of deep molecular models.

### 3. Handle uncertainty

Many critics over deep molecular models arise from the fact that assessment or validation of the models is very tricky. For example, we usually do not know and cannot tell when the model will perform well and when it will fail. Taking deep PES models as example, even if we get a "perfect" model which zeroes the error for the training data, we are still unsure about its performance in those regions uncovered by the training samples. This difficulty is caused by the deterministic learning mechanism adopted by current deep molecular models. Specifically, given an input **x**, the deterministic learning only predicts the value of the outcome $f_\theta(\mathbf{x})$, but leaves no information about the uncertainty of the prediction.

The lack of uncertainty assessment also calls trouble for the preparation and curation of the training data. For example, if we knew the prediction uncertainty, we would naturally prepare more training data over the regions with higher uncertainty. This procedure is known as active learning for data augmentation in machine learning,[106-107] because the data preparation is achieved through the user's active interactions with the model. Unfortunately, almost no existing deep molecular models provide the uncertainty information.

Currently, some researchers ameliorated this issue by training an ensemble of deep molecular models,[37, 61] so the uncertainty of predictions can be approximated by the divergence between the predictions of these models. However, this is not a theoretically justified approach, and a reasonable estimation of the uncertainty usually entails a large number of independent models. Considering that even a single deep molecular model would be computationally expensive, training an ensemble of models is often prohibitive in practice.

On the other hand, incorporation of uncertainty in statistical models is formally considered by Bayesian learning.[108] Particularly, Bayesian learning has given rise to many fruits for regression tasks which are ubiquitous in molecular modeling. For example, Gaussian Processes (GP) is famous for its ability to handle uncertainty in regression problems, and is widely used for functional regression or Bayesian optimization.[109] However, the vanilla GP is a non-parametric method so it cannot directly scale to high-dimensional input data. Nonetheless, there are increasing attempts to cross-fertilize Bayesian learning with deep learning.[110-111] We thus note here that augmenting the current deterministic learning mechanism with Bayesian learning may be a promising and reasonable direction for further investigation in order to handle the uncertainty of deep molecular models.

**4. Incorporate experiments**

Most of the existing applications of deep molecular modeling were dealing with data from simulations rather than experiments. However, we note there that deep molecular modeling can find wide applications in connection with experiments. In related studies, incorporating experimental information into molecular modeling and simulations is largely based on maximum entropy (MaxEnt) principle.[112-113] Although having been proved useful in many cases,[114-116] MaxEnt suffers some limitations. For example, MaxEnt may efficiently process low-dimensional restraints but becomes less feasible for inter-dependent high-dimensional ones. Furthermore, many single-molecule experiments provide information about the distribution of certain observables. However, MaxEnt can only incorporate the expectation value as restraints but wastes the information about the higher-order moments.

As an alternative approach, deep molecular modeling with EBMs treats the experimental observables as evidence so that one can implement the maximum likelihood estimation via Bayesian inference. For instance, incorporating experimental observations in EBMs is straightforward by virtue of the compositionality of the energy function,

$$F_\theta(\mathbf{x}) = F_{\text{exp}}\left(\mathbf{q}_{\text{exp}}(\mathbf{x})\right) + F_{\text{phys}}(\mathbf{x}) \tag{20}$$

where $F_{\text{exp}}$ corresponds to the experimentally measured quantities $\mathbf{q}_{\text{exp}}(\mathbf{x})$, whereas $F_{\text{phys}}$ accounts for additional physics restraints (e.g., force-field-like terms). Training of such an EBM will yield an energy function $F_\theta(\mathbf{x})$ whose marginal distribution over the experiment-related variables $\mathbf{q}_{\text{exp}}(\mathbf{x})$ is consistent with observations.[44] Specifically, both TALOS and VIFE represent available algorithms that can potentially achieve this goal. We thus expect deep molecular modeling with EBMs to find more applications in conjunction with experiments, especially single-molecule ones.

**5. End notes**

In addition to deep learning, non-parametric learning has also progressed significantly in the field of molecular modelling.[26, 117-118] Many of the above-mentioned limitations of deep molecular models can be readily circumvented or overcome by these non-parametric models. Therefore, the cross-fertilization between deep learning and non-parametric methods may lead to more reliable and efficient molecular models. Last but not the least, in order to democratize deep learning for molecular modeling and simulations, it is necessary to develop special-purpose hardware and software suitable for fast and user-friendly computation. For example, molecular simulation community will definitely benefit from the auto-differentiation and parallel computation which are the bedrocks of modern large-scale deep learning. Besides, as the machine learners and molecular simulation practitioners usually work in different platforms, a proper linker or interface which could unify the molecular modeling software and deep learning software is highly desired. Indeed, researchers from both scientific and business communities like Google are making efforts to this end.[119-120] With the improvement of the infrastructure, deep learning is expected to bring larger impact and more opportunities to molecular modeling and simulations in the near future.

**Acknowledgements**

The authors thank Weinan E, Wenjun Xie, Xing Che and Cheng-Wen Liu for useful discussion. This research was supported by National Natural Science Foundation of China [21927901, 21821004, 21873007 to Y.Q.G], the National Key Research and Development Program of China [2017YFA0204702 to Y.Q.G.] and Guangdong Basic and Applied Basic Research Foundation [2019A1515110278 to Y.I.Y.]. The author J.Z. thanks the Alexander von Humboldt Foundation for supporting part of this research.

**Notes:** The authors declare no conflict of interest.

# Tables and Figures

**Table 1.** Survey of deep molecular models

| Model | Type[a] | Birth Year[b] | Dataset[c] |
|---|---|---|---|
| BPNN[21] | Descriptor | 2007 | Silicon[21] |
| ANI[22] | Descriptor | 2017 | GDB[121] |
| MPNN[34] | GNN | 2017 | QM9[122] |
| DTNN[36] | GNN | 2017 | QM9 |
| TensorMol[23] | Descriptor | 2018 | Chemspider[23] |
| SchNet[35] | GNN | 2018 | MD17[118], QM9 |
| DPMD[2] | Descriptor | 2018 | MD17 |
| PhysNet[23, 37] | GNN | 2019 | MD17, QM9 |

a. Model types are either descriptor-based or GNN-based.
b. The year when the model was first proposed.
c. Dataset used for benchmarking the models.

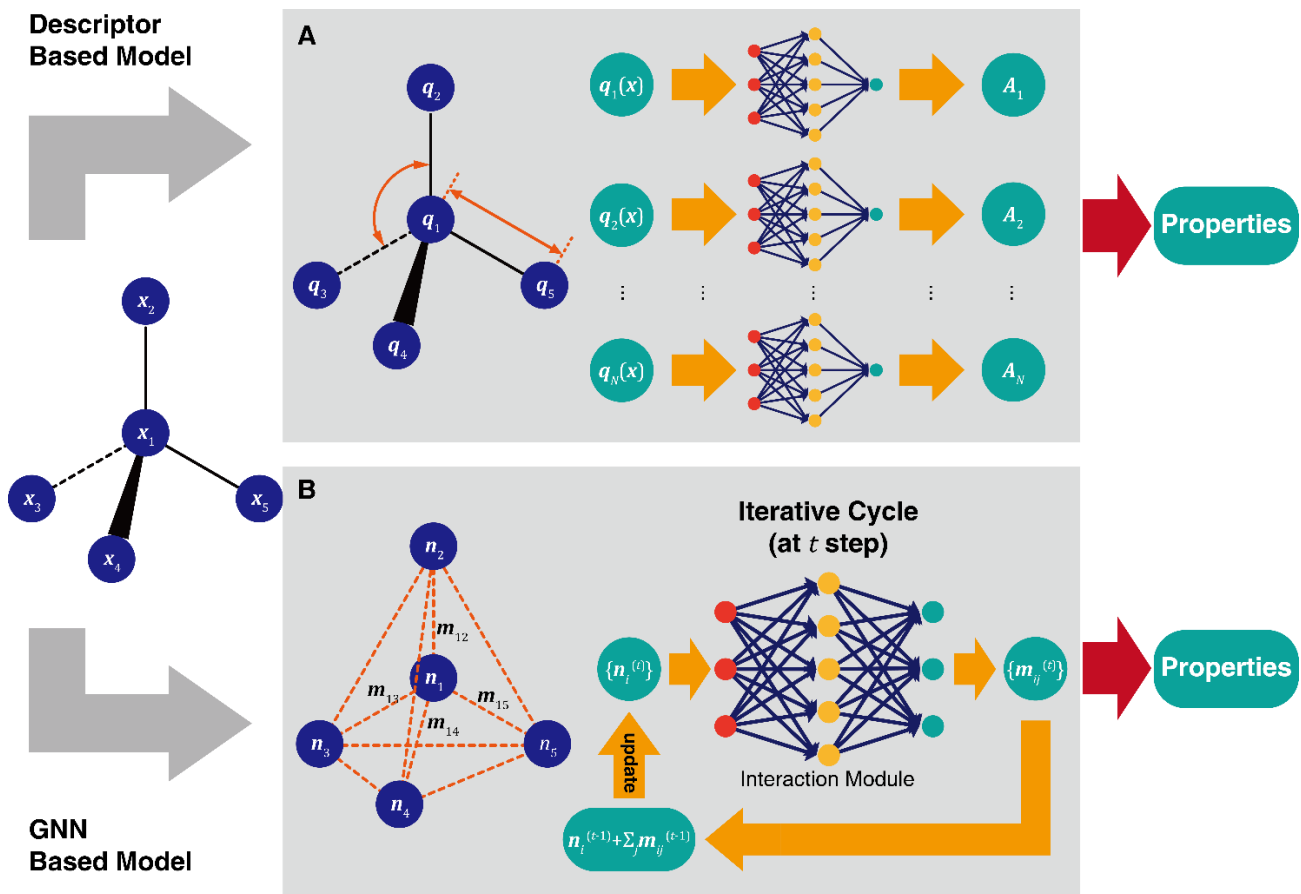

**Figure 1.** Architectures of deep molecular models. (A) Descriptor-based models (Eqs. (1-2)). $\mathbf{q}(\mathbf{x}_i)$ corresponds to the descriptors to the *i*-th particle (atom), which is further fed into a neural network. The output of each atomic neural network is then summed up to predict molecular properties. (B) GNN-based models (Eqs. (3-7)). Each particle (atom) is represented by a node vector ($\mathbf{n}_i$). An interaction module collects messages ($\mathbf{m}_{ij}$) between nodes which are used to update the node vector. After several iterations of message passing, the node vectors are used to predict the molecular properties.

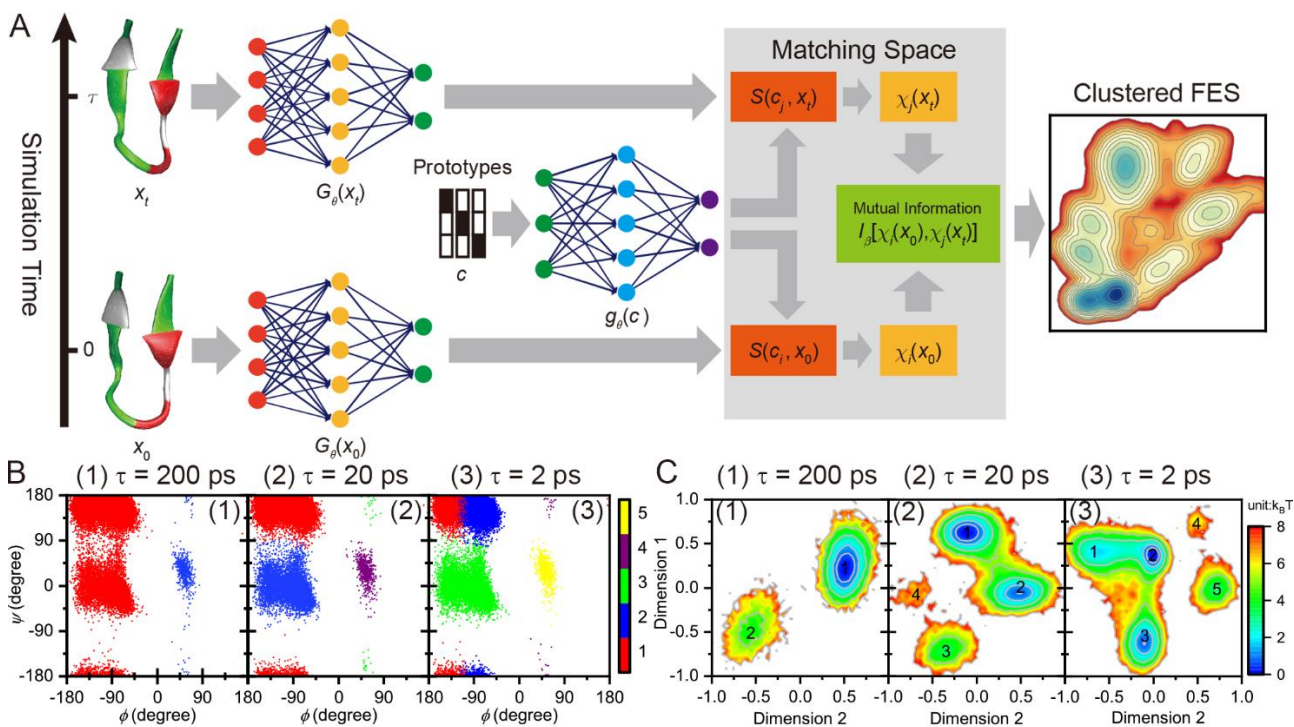

**Figure 2.** Information distilling of metastability (IDM). (A) Illustrative working flow of IDM. High-dimensional input vectors (**x**) and preimages of cluster centroids (**c**) are projected into images in the matching space by $G_\theta$ and $g_\theta$, respectively. Similarity score between **x** and **c** calculated in the matching space, $S(\mathbf{x}, \mathbf{c})$, leads to output $\chi(\mathbf{x})$ which is further used for Mutual Information ($I_\beta$) maximization. The matching space serves as the reduced embedding for **x**, and $\chi(\mathbf{x})$ as the soft clustering labels. (B) IDM clustering results for Ala2 over decreasing timescales. From left to right: (1) $\tau = 200$ ps, 2 clusters identified; (2) $\tau = 20$ ps, 4 clusters identified; (3) $\tau = 2$ ps, 5 clusters identified. Cluster identity is marked by different colors (with the color bar on the right). (C) IDM dimension reduction results for Ala2 over decreasing timescales (in the same order as in panel B). The PMF for each reduced representation is shown, and centers of the metastable states are indexed by numbers. Figures were reproduced with permission from Zhang et al., *J. Phys. Chem. Lett.* **2019**, *10* (18), 5571-5576. Copyright 2019 American Chemical Society.

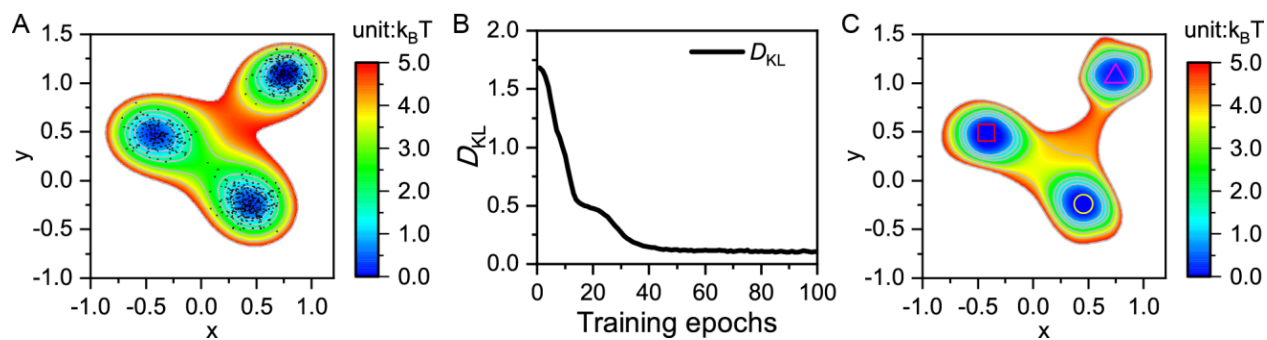

**Figure 3.** Application of variational inference of free energy (VIFE) on a 2D numerical potential. (A) 2D potential energy surface (PES), $U(x,y)$ of the model. Black dots correspond to representative samples randomly drawn from $p_{\text{FG}}$. (B) Evolution of the KL-divergence $D_{\text{KL}}(p_{\text{FG}}||p_\theta)$ against training epochs in VIFE. (C) The contour map corresponds to the optimized $F_\theta$ by VIFE, the three symbols (magenta triangle, red square and yellow circle) represent three free-energy minima found in $F_\theta$. Figures were reproduced with permission from Zhang et al., "Reinforcement Learning for Multi-Scale Molecular Modeling", e-print ChemRxiv: 9640814 (2019). Copyright 2019 Author(s), licensed under a CC BY-NC-ND 4.0 License.

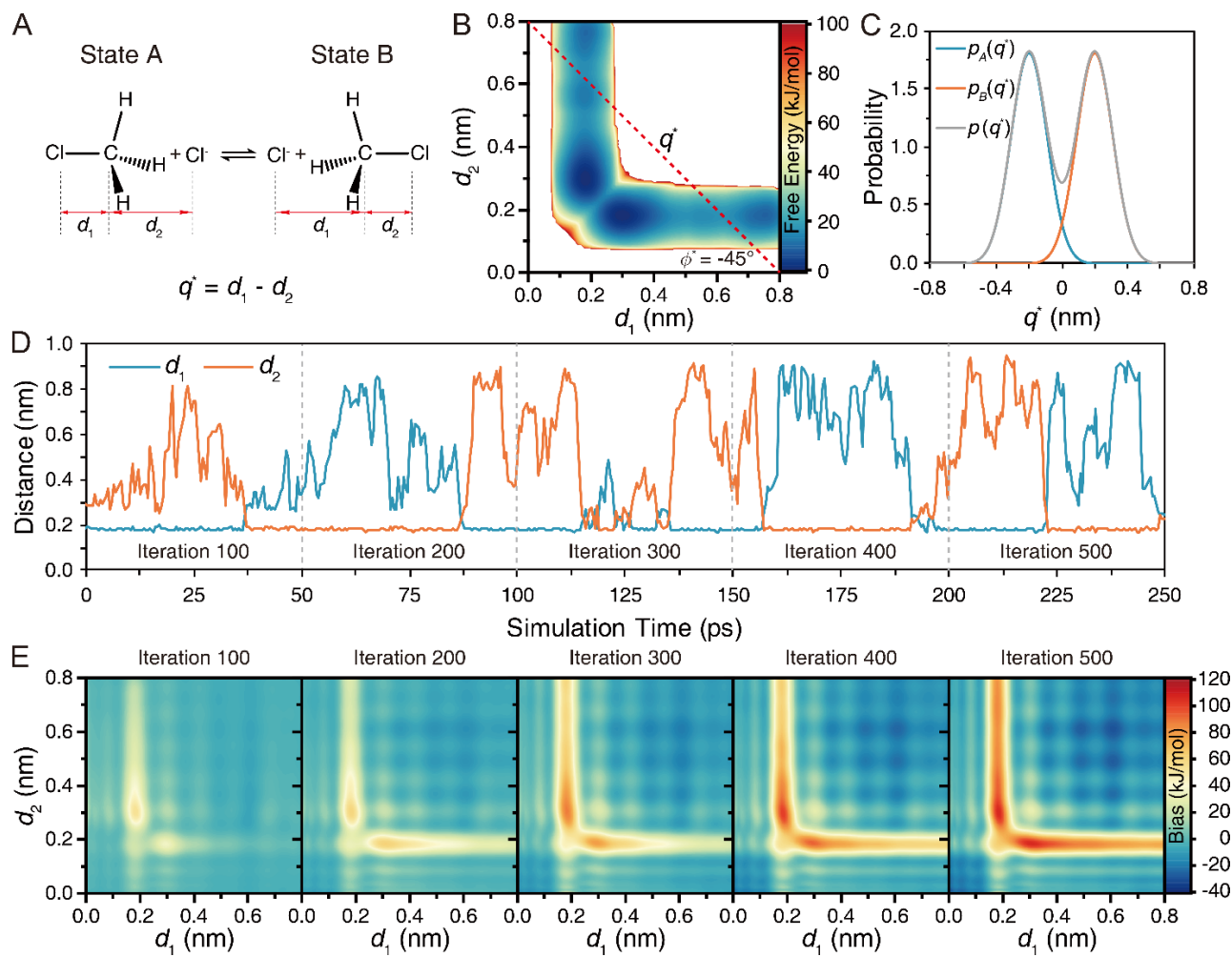

**Figure 4.** Targeted adversarial learning optimized sampling (TALOS) over a chemical reaction. (A) The reaction under study: $Cl^- + CH_3Cl \leftrightarrow CH_3Cl + Cl^-$. Two C-Cl bond distances $(d_1, d_2)$ are chosen as CVs, over which the bias potential is constructed; whereas a linear descriptor $q^* = d_1 - d_2$ is used to define the target distribution. (B) 2D FES over $(d_1, d_2)$ space and the illustration of $q^*$. (C) The distribution of $q^*$ for the reactant (blue), product (orange) and the interpolated target distribution (grey). (D) Examples of the biased MD trajectories projected on $d_1$ (blue) and $d_2$ (orange) at different TALOS training iterations. (E) The optimized 2D bias potential at different TALOS training iterations. Figures were reproduced with permission from Zhang et al., *J. Phys. Chem. Lett.* **2019**, *10* (19), 5791-5797. Copyright 2019 American Chemical Society.

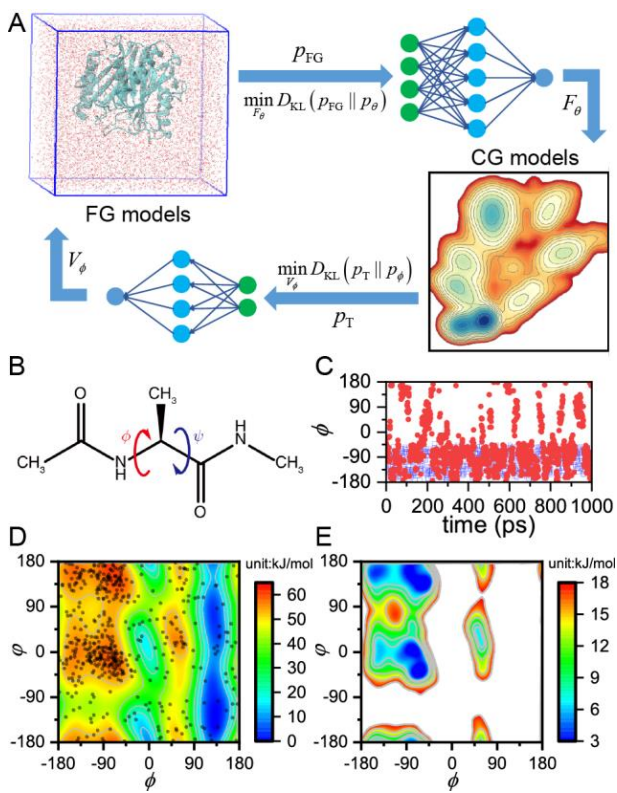

**Figure 5.** Reinforced variational inference of free energy (RE-VIFE). (A) Illustration of RE-VIFE. Given simulation samples from fine-grained (FG) models ($p_{FG}$), a coarse-grained (CG) potential $F_\theta$ can be variationally approximated. In turn, a target distribution $p_T$ can be defined based on CG simulations running over $F_\theta$, according to which a bias potential $V_\phi$ can be variationally optimized to boost the FG simulation. (B-E) RE-VIFE sampling of Ala2 in explicit water: (B) Structure of Ala2 and two CG variables are the torsions $\phi$ and $\varphi$. (C) 1-ns simulation trajectories projected on the slowly-changing torsion $\phi$. Blue squares correspond to vanilla MD, red dots to MD biased by $V_\phi$. (D) The contour map of $V_\phi$ optimized via RE-VIFE. Superimposed black dots are representative samples produced by the enhanced MD simulation under $V_\phi$. (E) Contour map of $F_\theta$ optimized via RE-VIFE. Figures were reproduced with permission from Zhang et al., "Reinforcement Learning for Multi-Scale Molecular Modeling", e-print ChemRxiv: 9640814 (2019). Copyright 2019 Author(s), licensed under a CC BY-NC-ND 4.0 License.